\documentclass[aps, prl, reprint, groupeaddress, floatfix]{revtex4-1}
\usepackage[T1]{fontenc}
\usepackage{amsmath}
\usepackage{amssymb}
\usepackage{epsfig}
\usepackage{graphicx}
\usepackage{braket}
\usepackage{color}
\begin{document}
\title{Demonstration of coherent time-frequency Schmidt mode selection using dispersion-engineered frequency conversion}

\author{Benjamin Brecht$^{1\star}$, Andreas Eckstein$^{1,2}$, Raimund Ricken$^1$, Viktor Quiring$^1$, Hubertus Suche$^1$, Linda Sansoni$^1$, and Christine Silberhorn${^1}$}

\affiliation{Integrated Quantum Optics, Applied Physics, University of Paderborn, Warburger Strasse 100 33098, Paderborn, Germany}	

\affiliation{Clarendon Laboratory, University of Oxford, Parks Road, Oxford OX1 3PU, United Kingdom}

\date{\today}

\begin{abstract}
	Time-frequency Schmidt (TFS) modes of ultrafast quantum states are naturally compatible with high bit-rate integrated quantum communication networks. Thus they offer an attractive alternative for the realization of high dimensional quantum optics. Here, we present a quantum pulse gate based on dispersion-engineered ultrafast frequency conversion in a nonlinear optical waveguide, which is a key element for harnessing the potential of TFS modes. We experimentally retrieve the modal spectral-temporal structure of our device and demonstrate a single-mode operation fidelity of 80\%, which is limited by experimental shortcomings. In addition, we retrieve a conversion efficiency of 87.7\% with a high signal-to-noise ratio of 8.8 when operating the quantum pulse gate at the single-photon level. 
\end{abstract}


\maketitle
The advent of the field of quantum information and computation has changed our way of thinking about information. The concept of mutually unbiased bases (MUBs) \cite{Schwinger:1960wq} lies at the heart of quantum information science applications like quantum key distribution \cite{Bennett:1984wv}, quantum state tomography \cite{Smithey:1993er} or entanglement detection \cite{Spengler:2012hy}. Most of these applications concentrate on two-dimensional systems, without exploiting the full potential of quantum mechanics. Only recently have people suggested utilizing higher-dimensional bases, which are of fundamental interest for questions addressing nonlocality \cite{Collins:2002bh,Vertesi:2010bq}. In addition, they provide larger alphabets, which promise increased security for quantum cryptography \cite{BechmannPasquinucci:2000ug,Cerf:2002fp}. A major requirement for any implementation of high-dimensional coding is a device which grants access to different basis states, in order to perform measurements in different MUBs. %

To date, the most widely used approach to exploit high-dimensionality is to deploy the orbital angular momentum (OAM) of photon pairs generated in parametric down-conversion (PDC) \cite{Mair:2001fd,Leach:2010tt,Dada:2011do}. Recent results demonstrated an increased information capacity \cite{Barreiro:2008jl} and increased security for quantum cryptography \cite{Groblacher:2006ec,Leach:2012gu}, as was predicted for high-dimensional coding. %

OAM states are appealing basis states, because they form a \textit{natural} basis for describing spatial entanglement in PDC \cite{Arnaut:2000kc,FrankeArnold:2002ee}. In addition, there exists an efficient mode sorter, which facilitates the deterministic separation of many OAM states using only linear optical elements \cite{Berkhout:2010cb}. On the downside, OAM states are incompatible with integrated single-mode network architectures, because they encode information in different spatial field modes. %
This directly implies, that OAM states cannot be generated with waveguided PDC sources, which feature high brightness and excellent compatibility with fiber networks \cite{Harder:2013hk}. %

However, PDC provides an alternative resource for high-dimensional information coding, namely energy-time entanglement \cite{Franson:1989uu,Kwiat:1993ew,Brougham:2014bn}. Here, the \textit{natural} basis functions are the so-called  time-frequency Schmidt (TFS) modes \cite{Law:2000wd}. %
Compared to OAM states, TFS modes offer three advantages for high-dimensional information coding: first, they are well-suited to integration, because they all live within the same spatial field distribution; second, a sophisticated toolbox for controlling the spectral-temporal structure of PDC exists \cite{URen:2005wb}, and results on waveguided PDC have already demonstrated the energy-efficient generation of single- and few-mode states with tunable spectral-temporal correlations \cite{Eckstein:2011wl,Harder:2013hk}; third, waveguided PDC guarantees an intrinsic control over the spatial degree of freedom, which is largely decoupled from the spectrum \cite{Mosley:2009kr}. %

The drawback of TFS modes is, that their manipulation cannot be accomplished with linear optical elements, and a mode sorter has not been available. %
As an answer to this, we have recently proposed a so-called quantum pulse gate (QPG) that is capable of selecting a single TFS mode from a high-dimensional input and convert it to a different frequency \cite{EcksteinA:2011vg, Brecht:2011hz}. Although theoretical studies suggest that the mode-selectivity of one single QPG is limited to around 87\% \cite{Christ:2013fg, Reddy:2013ip}, recent work from D. Reddy \textit{et al.} shows that this issue can be overcome \cite{Reddy:2014bt}.%

In this paper, we present the experimental implementation of a QPG: we demonstrate TFS single-mode operation with a fidelity of up to 80\%, and present a way of retrieving the QPG TFS mode structure in the experiment.  Moreover, we measure a conversion efficiency of close to 90\% when operating the device at the quantum level. %

The QPG is based on dispersion-engineered, ultrafast sum-frequency generation in a periodically poled lithium niobate waveguide. Due to the careful tailoring of the device, the input signal propagates through the waveguide at the same velocity as the pump pulses. %
In this case, the QPG operation on an input state $\ket{\psi}_\mathrm{in}$ is given by 
\begin{equation}
	\ket{\psi}_\mathrm{out} = \exp\left[\imath\theta\hat{A}\hat{C}^\dagger+\mathrm{h.c.}\right]\ket{\psi}_\mathrm{in},
	\label{eq:qpg}
\end{equation}
where the operators $\hat{A}$ and $\hat{C}^\dagger$ are TFS mode operators \cite{Law:2000wd}. They describe the annihilation of a photon in an ultrafast pulse with spectrum $\mathcal{A}(\omega)$ and the simultaneous generation of a photon in a pulse with spectrum $\mathcal{C}(\omega)$, which is centered at a different frequency. 

\begin{figure}
	\centering
	\includegraphics[width=.8\linewidth]{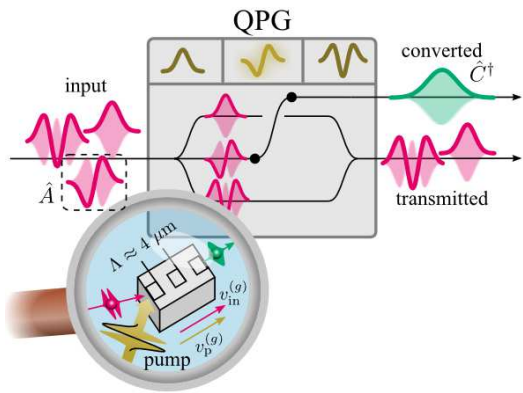}
	\caption{(Color online) Illustration of a QPG. A single TFS mode $\hat{A}$ from a multimode input state is selected and converted to an output mode $\hat{C}^{\dagger}$ at a different frequency. The remaining modes are simply transmitted. For more information see the text.}
	\label{fig:schematic}
\end{figure}

The working principle of a QPG is illustrated in Fig. \ref{fig:schematic}. An input state comprising multiple TFS modes (left) is sent into a QPG, which is adapted such that the highlighted TFS mode $\hat{A}$ is selected and converted into $\hat{C}^\dagger$. The remaining TFS modes of the input are transmitted. The magnified view shows the underlying device. Ultrafast, shaped pump pulses (lower left), which serve as a dial choosing the selected mode, are coupled to a periodically poled nonlinear waveguide, together with the input (upper left). Inside the waveguide, pump and input propagate at the same group velocity and a converted output is generated at the sum-frequency of input and pump. At the output of the QPG, converted and transmitted TFS modes can be separated using a dichroic mirror. Note that the shape of the selected TFS mode $\hat{A}$ is defined by the shape of the ultrafast pump pulses of the QPG, whereas the converted mode $\hat{C}^\dagger$ is given by the waveguide dispersion. %

The QPG can be interpreted as a special quantum mechanical beamsplitter, which operates on a single TFS mode of an ultrafast quantum state. The reflectivity or conversion efficiency of the QPG is given given by $\sin^2(\theta)$ (compare Eq. (\ref{eq:qpg})), where the coupling constant $\theta$ is a function of the pump power and complete conversion is, in principle, possible. %

The general idea to verify the QPG operation in the experiment, is to perform a tomographic reconstruction of the TFS mode characteristics of the QPG. In contrast to quantum process tomography, where photon-number properties of the process are evaluated \cite{Lobino:2008gg}, modal characteristics have the advantage that they are a mutual concept of classical and quantum light. Consequently they are accessible with \textit{classical} measurements only. This facilitates the complete characterization of the TFS mode structure by evaluating the impact of the QPG on a set of coherent probe states. We point out that probing one QPG with a set of probe states is tantamount to probing many different QPGs with one single probe state. 

Our coherent probe state exhibited a Gaussian TFS mode $\hat{G}^\dagger$, with an associated spectral distribution $\mathcal{G}(\omega)$. In addition, we implemented different QPGs by shaping of the classical pump pulses, thus changing the selected mode $\hat{A}$. Then, the QPG only selects the fraction of the probe that overlaps with $\hat{A}$. Consequently, the converted output intensity is proportional to $I_\mathrm{out}\propto|\int d\omega\,\mathcal{G}(\omega)\mathcal{A}^*(\omega)|$. Note that, although the measured quantity is an intensity, our approach is inherently phase-sensitive, since the overlap integral contains the complex-valued TFS mode spectra. In this way, it is possible to map the selected mode $\hat{A}$ by monitoring $I_\mathrm{out}$ for different realizations of the QPG. In contrast, the spectrum of the converted output is given by $\mathcal{C}(\omega)$ and thus grants direct access to the output mode $\hat{C}^\dagger$. 

\begin{figure}
	\centering
	\includegraphics[width=\linewidth]{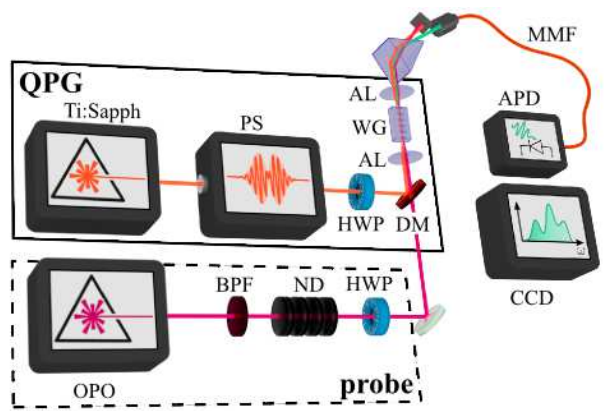}
	\caption{(Color online) Experimental setup. PS, pulse shaper; OPO, optical parametric oscillator; BPF, bandpass filter; ND, neutral density filter; HWP, half-wave plate; DM, dichroic mirror; AL, aspheric lens; WG, waveguide sample; MMF, multimode fiber; APD, avalanche photo diode; CCD, single-photon sensitive CCD spectrometer. For more information see the text.}
	\label{fig:setup}
\end{figure}

Our experimental setup is shown in Fig. \ref{fig:setup}. The probe state was prepared with an optical parametric oscillator (APE Compact OPO) that generated ultrafast pulses with a central wavelength of $1535\,$nm. These were subsequently filtered to a spectral bandwidth of $12\,$nm, corresponding to a pulse duration of $287\,$fs, and attenuated to the single-photon level. %
 The second part of the setup, the actual QPG, consists of a titanium sapphire oscillator (Coherent Chameleon Ultra II) generating $865\,$nm pulses with a maximum bandwidth of $7.9\,$nm, corresponding to a duration of $140\,$fs, which served as bright pump for the QPG. The pulses were sent through an acousto-optic pulse shaper (Fastlite Dazzler) to realize different pulse shapes and subsequently coupled to the waveguide sample. We deployed a homemade periodically poled waveguide with a remarkably low poling period of only $4.4\,\mu$m, which was temperature stabilised at $T=190^\circ$C to provide quasi-phasematching between the involved fields and at the same time minimize detrimental photorefractive effects. %
 Behind the waveguide, the converted $553\,$nm light was filtered and coupled into a multimode fiber, which was fed into either a single-photon sensitive CCD spectrometer (Andor iKon-M 934P-DD / Shamrock SR-303iA), or a silicon avalanche photo diode (Perkin Elmer SPCM-AQRH-13) for photon counting. 

\begin{figure}
	\centering
	\includegraphics[width=\linewidth]{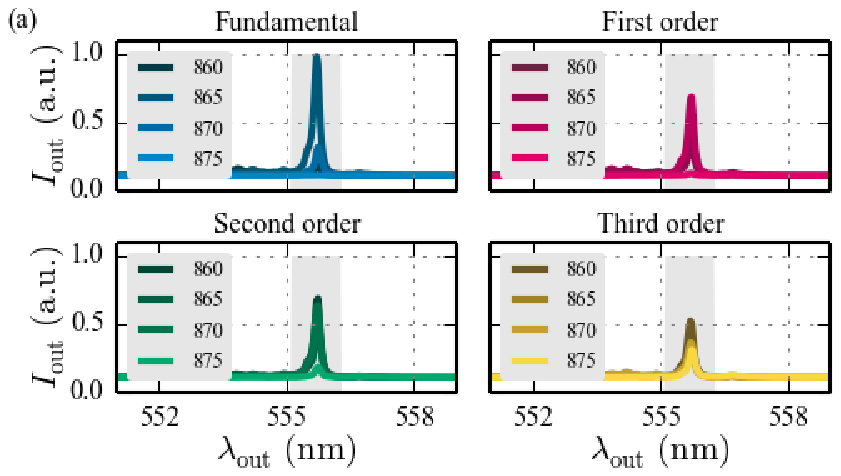}
	\includegraphics[width=\linewidth]{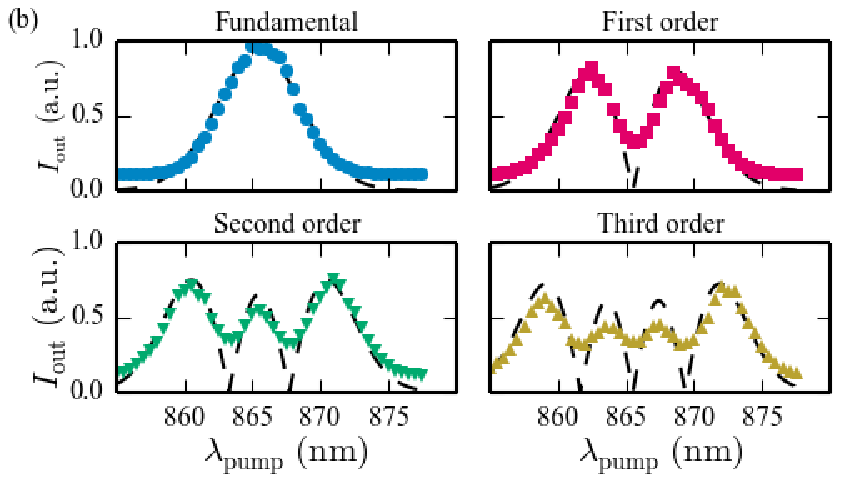}
	\caption{(Color online) Measurement results for the reconstruction of the QPG TFS modes. In (a), we plot chosen recorded output spectra for the indicated pump wavelengths, whereas in (b) we show the converted output intensity $I_\mathrm{out}$ as a function of the pump wavelength $\lambda_\mathrm{p}$. Note that error bars in (b) are smaller than the symbols. For more information see the text.}
	\label{fig:scanning}
\end{figure}

In the following, we present our measurement results. First, we characterise the QPG TFS mode structure, which at the same time facilitates the identification of the ideal QPG operation wavelength. Thereafter, we adapt the QPG bandwidth to the probe state. Having found the ideal operation point, we benchmark the QPG performance both in terms of mode selection and conversion efficiency at the single-photon level. %

Our first measurements disclose the QPG TFS mode structure. We utilize bright probe pulses with around 100 photons per pulse and realize different QPGs by shaping the pump pulses into four different Hermite-Gaussian modes with durations of $150\,$fs, which intentionally do not fit the probe state duration. The Hermite-Gaussian functions form a complete basis, in agreement with the Gaussian TFS mode $\mathcal{G}(\omega)$ of our probe state. For each QPG implementation, we scanned the central pump wavelength between $855\,$nm and $872\,$nm and recorded the converted output spectra with the CCD spectrometer.%

Some of these spectra are exemplarily shown in Fig. \ref{fig:scanning}(a). From top left to bottom right, the basic TFS shape of the pump changes from a fundamental Gaussian to a third-order Hermite-Gaussian. From our theory we expect that the output TFS mode $\hat{C}^\dagger$ of the QPG is only defined by the phasematching of the waveguide. This is verified by the measured spectra, which are similar regardless of the spectral-temporal pump shape. For applications this means, that formerly orthogonal TFS modes can be interfered after the QPG operation. %

From the measured spectra we retrieve a spectral bandwidth of $\Delta\lambda_\mathrm{out}\approx0.14\,$nm, corresponding to a pulse duration of $3.2\,$ps, which demonstrates a bandwidth compression about a factor of eleven. This factor, which is of interest for applications aiming for interfacing flying and stationary qubits with vastly different bandwidths, can be increased when deploying longer waveguides.  %

In Fig. \ref{fig:scanning}(b), we plot the output intensities $I_\mathrm{out}$ as a function of the pump wavelength $\lambda_\mathrm{p}$. The intensities were calculated from the spectra by integrating over the grey-shaded area in Fig. \ref{fig:scanning}(a). From top left to bottom right, the spectral-temporal pump shape again changes from a fundamental Gaussian to a third-order Hermite-Gaussian. We recall that $I_\mathrm{out}\propto|\int d\omega\,\mathcal{G}(\omega)\mathcal{A}^*(\omega)|$, and find that the measured curves nicely reproduce the pump shape, indicated by the dashed black lines in the respective diagrams. Deviations occur only in the regions of sharp features, which we can attribute to the limited resolution of the pulse shaper of roughly $0.7\,$nm. This measurement demonstrates that the input mode $\hat{A}$ is defined by the shape of the ultrafast pump pulses, as was expected from theory (see also the Supplemental Material \cite{Supplements}). %

We also identify an ideal central pump wavelength at $\lambda_\mathrm{p}=865.6\,$nm, where the conversion for the Gaussian is highest, and similarly the conversion for the first-order mode exhibits a minimum. Next, we demonstrate the second step towards optimization of $\hat{A}$ with respect to the probe state. We fix the central pump wavelength at the optimal value of $\lambda_\mathrm{p}=865.6\,$nm and change the spectral pump bandwidth $\Delta\lambda_\mathrm{p}$. Again, we record the output intensity for the different spectral-temporal pump shapes from Fig. \ref{fig:scanning}. 

\begin{figure}
	\centering
	\includegraphics[width=\linewidth]{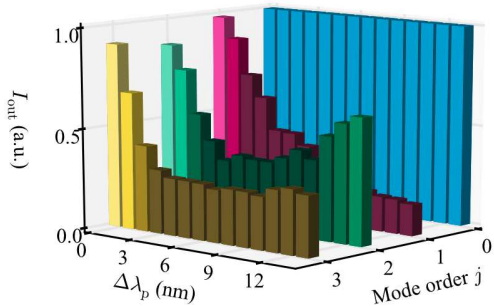}
	\caption{(Color online) Measurement results for the adaption of the QPG TFS modes to an unknown input. For more details see the text.}
	\label{fig:adaption}
\end{figure}

Our measurement results are shown in Fig. \ref{fig:adaption}, where we plot the normalized output intensity $I_\mathrm{out}$ against the pump bandwidth, where normalization was required to account for the imperfections of the pulse shaper. As expected for Hermite-Gaussian modes, the conversion efficiency for the odd order pump modes drops to a minimum after the resolution limit of the pulse shaper is overcome, and does not increase again.  In contrast, the conversion efficiency for the second-order pump mode decreases to a minimum and then starts to increase again, when further increasing $\Delta\lambda_\mathrm{p}$. From the measurements, we deduce from this minimum an \textit{ideal} pump bandwidth of $\Delta\lambda_\mathrm{p}\approx4.0\,$nm, corresponding to a pump duration of around $275\,$fs. This is in excellent agreement with the duration of the probe pulses of $287\,$fs, as we would have expected. %

\begin{figure}
	\centering
	\includegraphics[width=.8\linewidth]{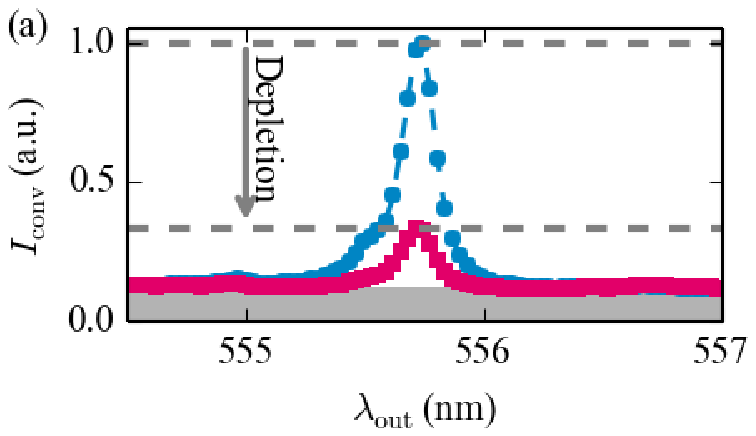}
	\includegraphics[width=.8\linewidth]{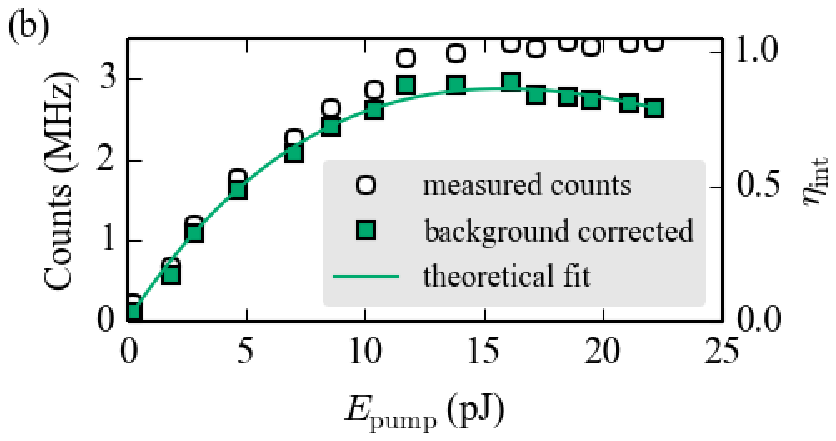}
	\caption{(Color online) Performance benchmarks of our QPG. From the spectral suppression in (a), we obtain a mode-selectivity of 80\%, whereas the efficiency measurement in (b) reveals an internal conversion efficiency of 87.7\%. In both plots, error bars are smaller than the symbols. For more information see the text.}
	\label{fig:benchmark}
\end{figure}

Now, we benchmark the performance of the QPG. To this end, we fix the pump parameters at the optimized values of $\lambda_\mathrm{p}=865.6\,$nm and $\Delta\lambda_\mathrm{p}=4.0\,$nm. Then, we switch the pump mode from Gaussian to first-order Hermite-Gaussian and record the converted output spectra. These are shown in Fig. \ref{fig:benchmark} (a). The blue dotted spectrum corresponds to a Gaussian pump, whereas the red squared spectrum was taken with a Hermite-Gaussian pump, respectively. When subtracting the flat spectral background caused by the spectrometer noise, marked as a grey area, we obtain a depletion or \textit{mode-selectivity} \cite{Reddy:2013ip} of 80\%, which, in this experiment, is limited by the finite resolution of the pulse shaper. %

Finally, we investigate the noise performance and conversion efficiency of the QPG when operated at the single photon level. Therefore, we attenuate the probe states to a mean photon number of $\langle n\rangle\approx0.15\,$photons/pulse and record the converted output counts with the APD. In Fig. \ref{fig:benchmark} (b), we plot the recorded counts versus the pump pulse energy. We retrieve the background by blocking the probe state and recording the remaining counts for each pump energy. The corrected counts (green squares) are in excellent agreement with the theoretical $\sin^2$-fit. In addition, the signal-to-noise ratio for maximum conversion is roughly 8.8, which demonstrates a low-noise operation of the QPG required for quantum applications. We deduce an internal conversion efficiency, defined as the number of converted photons at the end of the waveguide versus the number of probe photons at its input, of 87.7\%, where we did not correct for waveguide propagation losses. This maximum is reached at a pump pulse energy of only $E_\mathrm{p}\approx16\,$pJ in front of the waveguide incoupling, corresponding to a cw-equivalent power as low as $1.3\,$mW for an $80\,$MHz repetition rate. This remarkable efficiency is possible only due to the careful tailoring of our device. Since pump and probe pulses propagate through the waveguide at the same velocity, the probe always experiences the high peak power of the pump pulses. In addition, the TFS single-mode behavior facilitates an extra-ordinary energy efficiency, which becomes a key feature when considering large-scale quantum networks where energy consumption becomes crucial. 

In conclusion, we implemented a QPG and retrieved its TFS mode characteristics by probing it with coherent probe states. Our device facilitates a single-mode operation on the TFS modes of ultrafast quantum states, with a mode-selectivity of at least 80\%, limited mainly by the finite resolution of our pulse shaper. This operation facilitates measurements in different, high-dimensional mutually unbiased spectral-temporal bases, which is a prerequisite for many quantum information applications, for instance quantum cryptography with increased security \cite{Groblacher:2006ec,Leach:2012gu}. In addition, our device provides a high internal conversion efficiency of 87.7\% and a good signal-to-noise ratio of 8.8 when operated at the single photon level. Since the output mode of the QPG is independent of the selected mode, the QPG facilitates interfacing between orthogonal TFS modes. Moreover, when operated at low conversion efficiencies, the QPG can be exploited for the implementation of TFS mode-selective non-Gaussian operations in multimode continuous-variable quantum information schemes \cite{Averchenko:2014dn}. We expect our work to impact a wide range of applications in discrete and continuous variable quantum information. 

The authors acknowledge helpful discussion with Andreas Christ, Fabian Katzschmann, and Michael Stefszky. This work has been supported by the EC via QESSENCE and by the Deutsche Forschungsgemeinschaft (DFG) via TRR142. %

%
%
\bibliography{bib}

\end{document}